\documentclass{article}
\usepackage[dvips]{graphicx}
\begin{document}
\title{Stochastic synchronization in globally coupled phase oscillators}
\author{Hidetsugu  Sakaguchi\\
Department of Applied Science for Electronics and Materials,\\ Interdisciplinary Graduate School of Engineering Sciences,\\
Kyushu University, Kasuga, Fukuoka 816-8580, Japan}
\maketitle
\begin{abstract}
Cooperative effects of periodic force and noise in globally 
coupled systems are studied using a nonlinear diffusion equation for the number density. 
 The amplitude of the order parameter oscillation is enhanced in an intermediate range of noise strength for a globally coupled bistable system, and the order parameter oscillation is entrained to the external periodic force in an intermediate range of  noise strength. These enhancement phenomena of the response of the order parameter in the deterministic equations are interpreted as stochastic resonance and stochastic synchronization in globally coupled systems.\\
\\
PACS: 05.40.+j, 05.70.Ln, 02.50-r\\
\end{abstract}
\section{Introduction}
Recently, various noise effects for nonlinear systems have been studied.
In stochastic resonance, the response of a bistable system or an excitable system to a periodic force is enhanced with the addition of noise.
The stochastic resonance improves signal detection by the superposed noise \cite{rf:1,rf:2,rf:3}.  Noise-enhanced frequency locking is observed in stochastic bistable systems driven by a relatively strong periodic force or a chaotic signal and phase diagrams similar to the Arnold tongues are obtained \cite{rf:4,rf:5,rf:6}. 
 Noise induced entrainment among coupled oscillators is found experimentally in   Belousov-Zhabotinsky reactions and brain waves \cite{rf:7,rf:8}.
Stochastic resonance has been studied also in globally or locally coupled systems of many bistable elements and it was shown that the coupling can lead to the enhancement of the response \cite{rf:9,rf:10,rf:11,rf:12}. 
Various types of Fokker-Planck equations have been used to study the stochastic resonance  theoretically. 
 
On the other hand, various types of collective dynamics in globally coupled oscillators have been studied \cite{rf:13,rf:14}.  
Globally coupled phase oscillators under external noises can be studied with a nonlinear diffusion equation for the number density. 
The nonlinear diffusion equation is obtained from the mean field approximation of the Fokker-Planck equation for the $N$ oscillators, which is considered to be correct in the globally coupled system in the limit $N\rightarrow \infty$ \cite{rf:13}.
The nonlinear diffusion equation for the globally coupled phase oscillators 
can be transformed into  coupled nonlinear equations for the Fourier amplitudes of the number density. 
The numerical simulation of the coupled nonlinear equations for the Fourier modes is relatively easy.
Various nonequilibrium phase transitions were found in the globally coupled phase oscillators using the numerical simulation of the coupled equations for the Fourier modes and their bifurcation analysis \cite{rf:15}.

In this paper, we study an extended model of the coupled phase oscillators, in which an external periodic force is added to the model studied in [15]. 
Each element may exhibit stochastic resonance and stochastic synchronization, however, we study dynamical behaviors of the order parameter.
 The nonlinear diffusion equation is a deterministic equation, and the 
stochastic behaviors for each element are averaged out in the description of the order parameter.
The resonant response and the synchronization to a periodic force are more clearly shown in this deterministic system.
\section{Stochastic resonance in globally coupled bistable systems} 
At first, we consider a globally coupled bistable system. 
Each phase oscillator evolves according to an equation
\begin{equation}
\frac{d\phi}{dt}=-b\sin2 \phi-c\sin \omega_0t\sin\phi,
\end{equation}
where $\phi$ is the phase of the oscillator, $b$ and $c$ are positive constants, and  $\omega_0$ is the frequency of the periodic  force. 
There are two stationary solutions: $\phi=0$ and $\pi$. If the coefficient of the second term on the right-hand side  is not temporally periodic as $c\sin \omega_0 t$ but constant $c_0$,  the two solutions are bistable for $|c_0|\le 2b$.  In that case, the dynamical system has a potential function $U(\phi)=-b/2\cos 2\phi-c_0\cos\phi$, and the solution $\phi=0$ has a lower potential for $c_0> 0$ and  a higher potential for $c_0< 0$ than the solution $\phi=\pi$.
A model of a globally coupled system with a noise term is written as
\begin{equation}
\frac{d\phi_i}{dt}=-b\sin 2\phi_i-c\sin \omega_0 t\sin \phi_i-\frac{K}{N}\sum_{j=1}^{N}\sin(\phi_i-\phi_j)+\xi_i,
\end{equation}
where $N$ is the total number of elements, and the second term on the right-hand side represents the periodically forcing term and the third term represents mutual coupling, and the last term represents Gaussian white noise characterized by
\[\langle \xi_i(t)\rangle=0,\,\,\langle \xi_i(t)\xi_j(t^{\prime})\rangle=2D\delta_{ij}\delta(t-t^{\prime}),\]
where $D$ represents noise strength.
The mean field treatment holds exactly in the limit of $N\rightarrow \infty$. 
The normalized number density defined by
\[n(\phi,t)=\frac{1}{N}\sum_m\sum_j\delta(\phi_j-\phi-2\pi m)\]
obeys a nonlinear diffusion equation
\begin{eqnarray}
\frac{\partial}{\partial t}n(\phi,t)&=&-\frac{\partial}{\partial \phi}\left \{-b\sin 2\phi-c\sin\omega_0 t\sin \phi-K\int_{-\pi}^{\pi}d\phi^{\prime}\sin(\phi-\phi^{\prime})n(\phi^{\prime},t)\right\}n(\phi,t)\nonumber\\
& &+D\frac{\partial^2}{\partial \phi^2}n(\phi,t)
\end{eqnarray}
The $2\pi$-periodic function $n(\phi,t)$ can be expanded as
\[n(\phi,t)=\frac{1}{2\pi}\sum_{m=-\infty}^{\infty}\rho_m(t)e^{im\phi}\]
The nonlinear equation (3) is rewritten with coupled ordinary differential equations,
\begin{equation}
\frac{d\rho_m}{dt}=m\{K/2(\rho_1\rho_{m-1}-\rho_{-1}\rho_{m+1})+b/2(\rho_{m-2}-\rho_{m+2})+c/2\sin\omega_0 t(\rho_{m-1}-\rho_{m+1})\}-Dm^2\rho_m.
\end{equation} 
We have performed numerical simulation of Eq.~(4) retaining the first 50 modes with  the Runge-Kutta method of timestep $\Delta t=0.0001$.
The order parameter is expressed as 
\[\sigma=(1/N)\sum_{j=1}^Ne^{i\phi_j}=\int_{-\pi}^{\pi}d\phi e^{i\phi}n(\phi,t)=\rho_{-1}(t)=\rho_1^{*}.\]
\begin{figure}[htb]
\begin{center}
\includegraphics[width=12cm]{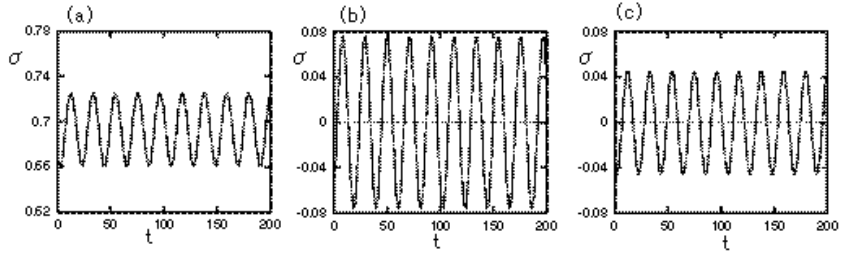}
\caption{Time evolutions of $\sigma(t)$ by Eq.~(4) at $D=0.4$ (b) $D=0.57$ and (c) $D=1$ for $b=0.2,\,c=0.05,\,\omega_0=0.3$ and $K=1$.   
} 
\label{fig:1} 
\end{center}
\end{figure} 
In this model, the order parameter takes a real value, that is, ${\rm Im}\rho_1(t)=0$. Figure 1 displays three time evolutions of $\sigma(t)$ at $D=0.4, 0.57$ and 1 for $b=0.2,\,c=0.05,\,\omega_0=0.3$ and $K=1$. The order parameter is regularly oscillating with the frequency $\omega_0$ of the external force. The average value of the order parameter takes a nonzero value at $D=0.4$ and zero at $D=0.57$ and 1.
The amplitude of the periodic oscillation is maximum at $D=0.57$.
We have calculated the average values $\langle \sigma\rangle$ of the order parameter and the temporal 
fluctuations around the average value, $\langle (\delta \sigma(t))^2\rangle^{1/2}$, where $\delta \sigma(t)=\sigma(t)-\langle \sigma\rangle$. The results are shown in Fig.~2. The average values of order parameters take nonzero values for $D <D_c\sim 0.56$. This is a kind of symmetry breaking phase transition induced by noises.  The amplitude of the periodic oscillation of the order parameter takes a maximum near the phase transition point.  The response function to the periodic modulation becomes large  
in an intermediate range of noise strength. This is interpreted as a kind of stochastic resonance in this globally coupled bistable system. 
A phase transition in a globally coupled bistable system without a periodic force and the enhancement of the susceptibility to a weak periodic force was discussed in   [9],[10]. 
\begin{figure}[htb]
\begin{center}
\includegraphics[width=11cm]{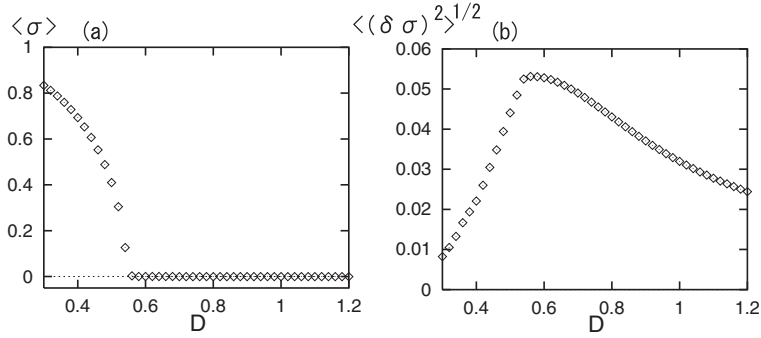}
\caption{(a) Average value $\langle \sigma\rangle$ of the order parameter  
and (b) the temporal 
fluctuations $\langle (\delta \sigma)^2\rangle^{1/2}$ as a function of $D$ for $b=0.2,\,c=0.05,\,\omega_0=0.3$ and $K=1$.
} 
\label{fig:2} 
\end{center}
\end{figure} 
We have found that a kind of phase transition occurs even for nonzero amplitude of periodic modulation.  

\section{Stochastic synchronization in globally coupled phase oscillators} 
Next, we consider a globally coupled oscillator system. 
Each phase oscillator is assumed to obey an equation
\begin{equation}
\frac{d\phi}{dt}=\omega-(b+c\sin\omega_0t)\sin\phi,
\end{equation}
where $\phi$ is the phase of the oscillator, $\omega,b$ and $c$ are positive constants, and  $\omega_0$ is the frequency of the periodic  force. 
If $c=b=0$, this equation describes a simple phase rotator, since the solution is $\phi=\omega t$, where $\omega$ is the natural frequency.
If $c=0$, and $b\ne 0$ but $b<\omega$, $d\phi/dt$ is not constant but always takes a positive value, and this equation describes still a phase rotator.
The natural frequency of the oscillator is given by $\sqrt{\omega^2-b^2}$. 
When $c=0$ and $b<\omega$ but $b$ is close to $\omega$, the oscillator behaves like the relaxation oscillation (it may be suitable to call it relaxation oscillation of phase), since the phase motion is clearly separated into fast motion and very slow motion near $\phi=\pi/2+2n\pi$ ($n$ is an integer). 
On the other hand, when $\omega < b$ and $c=0$,  the equation has a stable stationary solution given by $\phi=\sin^{-1}(\omega/b)$ and all trajectories are attracted to the stable point. 
If $c=0$, and $b>\omega$ but $b$ is close to $\omega$, the system behaves like an excitable system, since one-cycle rotation is excited by some small perturbation to the stationary state. 
The parameter $b$ is a parameter which determines a transition from an excitable system to an oscillatory system.  The threshold value is $b=b_c=\omega$. 
The parametric perturbation $b+c\sin\omega_0t$ implies that the threshold value is periodically modulated. 
It is known in an experiment of light-sensitive Belousov-Zhabotinsky reaction that illumination can control the threshold . The noise entrainment was observed by controlling the illumination \cite{rf:7}. The periodic modulation in our model may be related to such periodic modulation of the threshold value in experimental systems. 

A model of a globally coupled system with a noise term is written as
\begin{equation}
\frac{d\phi_i}{dt}=\omega-(b+c\sin\omega_0t)\sin\phi_i-\frac{K}{N}\sum_{j=1}^{N}\sin(\phi_i-\phi_j)+\xi_i,
\end{equation}
where $N$ is the total number of elements, and the third term represents mutual coupling, and the last term represents Gaussian white noise characterized by
\[\langle \xi_i(t)\rangle=0,\,\,\langle \xi_i(t)\xi_j(t^{\prime})\rangle=2D\delta_{ij}\delta(t-t^{\prime}).\]
A coupled phase oscillator model similar to this equation was studied by Kim et al. \cite{rf:16}. In their model, time delay was further assumed. In their system with time delay, there are multistable states, i.e., fast and slow synchronized states and a desynchronized state. They studied noise induced transitions among the multistable states in one oscillator system and a coupled system of 10 oscillators. We do not consider such time delay effect and study the stochastic synchronization using a nonlinear diffusion equation for a coupled system in the limit $N\rightarrow\infty$. 
The normalized number density 
obeys 
\begin{eqnarray}
\frac{\partial}{\partial t}n(\phi,t)&=&-\frac{\partial}{\partial \phi}\left \{\omega-(b+c\sin\omega_0 t)\sin \phi-K\int_{-\pi}^{\pi}d\phi^{\prime}\sin(\phi-\phi^{\prime})n(\phi^{\prime},t)\right\}n(\phi,t)\nonumber\\
& &+D\frac{\partial^2}{\partial \phi^2}n(\phi,t)
\end{eqnarray}
The corresponding coupled ordinary differential equations are written as
\begin{equation}
\frac{d\rho_m}{dt}=m\{K/2(\rho_1\rho_{m-1}-\rho_{-1}\rho_{m+1})+(b+c\sin\omega_0t)/2(\rho_{m-1}-\rho_{m+1})\}-(im\omega+Dm^2)\rho_m.
\end{equation} 
We have performed numerical simulation of Eq.~(8) retaining the first 50 modes 
with the Runge-Kutta method of timestep $\Delta t=0.0005$.
In this model, the order parameter $\sigma=\rho_{-1}$ is a complex variable. 
\begin{figure}[htb]
\begin{center}
\includegraphics[width=13cm]{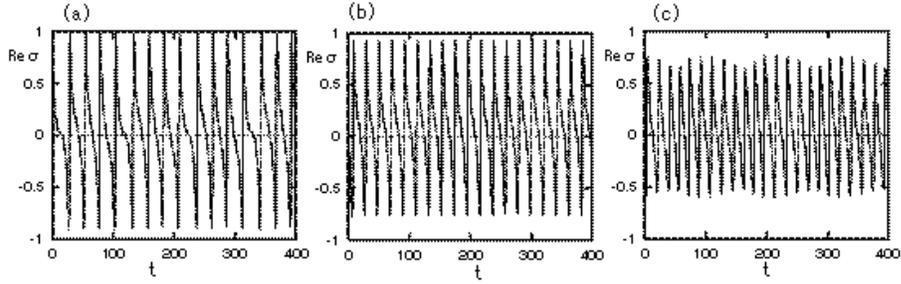}
\caption{Time evolutions of $\sigma(t)$ by Eq.~(8) at $D=0.04$ (b) $D=0.08$ and (c) $D=0.12$ for $b=1,\,c=0.02,\,\omega=1.005,\,\omega_0=0.3$ and $K=1$.   
} 
\label{fig:3} 
\end{center}
\end{figure} 
Figure 3 displays three time evolutions of ${\rm Re} \sigma(t)$ at $D=0.04,\,0.08$ and 0.12 for $\omega=1.005,\,K=1,\,\omega_0=0.3$ and $c=0.02$. 
The order parameter is entrained to the external periodic force at $D=0.08$. However, the motion of the order parameter is not completely entrained to the periodic force at $D=0.04$ and 0.12, and quasi-periodic motions appear.
\begin{figure}[htb]
\begin{center}
\includegraphics[width=12cm]{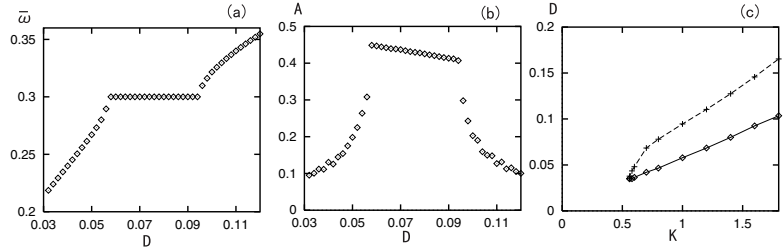}
\caption{(a) Frequency $\bar{\omega}$ of the order parameter oscillation vs. $D$ for $b=1,\,c=0.02,\,\omega=1.005,\,\omega_0=0.3$ and $K=1$.
(b) Fourier amplitude $A=(1/T)|\int_0^Tdt\sigma(t)\exp(-i\omega_0t)|$ vs. $D$.
(c) Phase diagram for the entrainment in $K-D$ plane. The entrainment is observed in the parameter range surrounded by two curves.
} 
\label{fig:4} 
\end{center}
\end{figure} 
\begin{figure}[htb]
\begin{center}
\includegraphics[width=11cm]{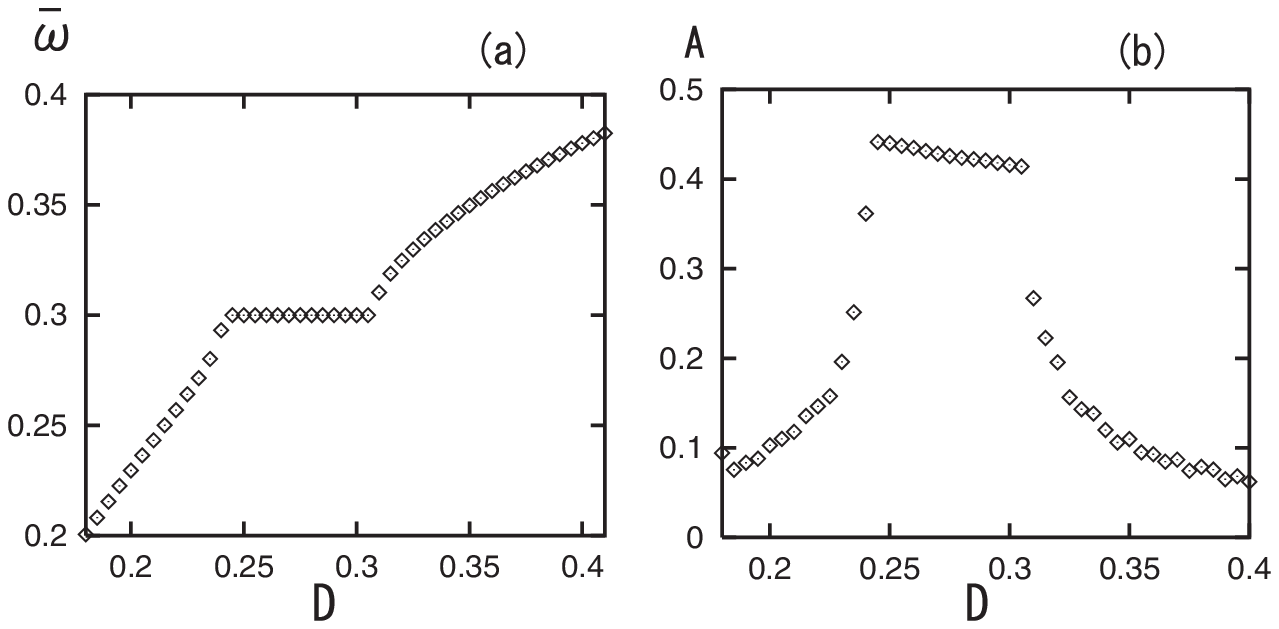}
\caption{(a) Frequency $\bar{\omega}$ of the order parameter oscillation vs. $D$ and 
(b) Fourier amplitude $A=(1/T)|\int_0^Tdt\sigma(t)\exp(-i\omega_0t)|$ vs. $D$   for $b=1,\,c=0.02,\,\omega=0.97,\,\omega_0=0.3$ and $K=2$.
} 
\label{fig:5} 
\end{center}
\end{figure} 
Figure 4(a) displays the frequency $\bar{\omega}$ of the motion of the order parameter as a function of $D$  for $\omega=1.005,\,K=1,\,b=1\,\omega_0=0.3$ and $c=0.02$. The frequency was calculated from the rotation number of $({\rm Re} \sigma(t),{\rm Im}\sigma(t))$ around the origin $(0,0)$. 
The frequency of the order parameter increases with the noise strength $D$.
It can be understood from the dynamical behavior of each oscillator.  
Each oscillator behaves like the relaxation oscillation and the phase motion becomes slow near $\phi=\pi/2+2n\pi$ when $D=0$, since $\omega\sim b$.  An effect of noises 
 is to make the regular slow motion randomly faster, and then the average frequency increases with the noise strength.  
The periodic modulation is further applied. 
The frequency entrainment is observed between $0.058<D<0.094$. This is a kind of the stochastic synchronization, although the motion of the order parameter is not stochastic. 
Figure 4(b) displays the Fourier amplitude of the order parameter motion with frequency $\omega_0$, that is, $(1/T)|\int_0^Tdt\sigma(t)\exp(-i\omega_0t)|$. 
The Fourier amplitude takes a maximum in the parameter range where the stochastic synchronization is observed. 
Figure 4(c) displays a parameter region in the $K-D$ parameter space, in which the synchronization is observed.  
This phase diagram seems to be the Arnold tongues discussed in [4-6], however, the ordinate is not the amplitude $c$ of the periodic forcing but the coupling constant $K$, so the meaning is different. 
For a fixed value of noise strength, the coupling constant also needs to take an intermediate  value for the stochastic synchronization to appear, as shown in this figure. Too large coupling constant makes the distribution of the number density too narrow,  and noise effects are effectively reduced.  

Even for a parameter range $\omega<b-c$, in which each elemental system is purely excitable without noises,  noises can induce phase slips and the oscillation is observed on an average.  When the periodic force is further applied, the stochastic synchronization can occur.  We have performed a numerical simulation for $\omega=0.97,\,b=1,\,K=2,\,\omega_0=0.3$ and $c=0.02$. 
Figure 5(a) displays the frequency of the motion of the order parameter as a function of $D$. The frequency entrainment is observed for $0.245<D<0.305$.  
Figure 5(b) displays the Fourier amplitude of the order parameter motion with frequency $\omega_0=0.3$. The Fourier amplitude also has  a maximum in an intermediate value of noise strength. 
\section{Summary}
We have proposed globally coupled phase oscillator models and performed 
numerical simulations of the corresponding nonlinear diffusion equations for the number density.  The response to the periodic force becomes large in an intermediate range of noise strength. It can be interpreted as a stochastic resonance or a stochastic synchronization, although the motion of the order parameter is regular.  We have shown numerical results for parametrically perturbed system, but we have also studied an additively perturbed model: 
\[\frac{d\phi_i}{dt}=\omega-b\sin\phi_i+c\sin\omega_0t-\frac{K}{N}\sum_{j=1}^{N}\sin(\phi_i-\phi_j)+\xi_i,\]
and another type of forced model:
\[\frac{d\phi_i}{dt}=\omega-b\sin\phi_i-c\sin(\phi_i-\omega_0t)-\frac{K}{N}\sum_{j=1}^{N}\sin(\phi_i-\phi_j)+\xi_i.\]
We have observed similar stochastic synchronization also in these systems.
It does not seem to depend on the detailed form of periodic forcing.
The stochastic synchronization may be found in many systems driven by noises and periodic forcing, if the parameter for each element is near a transition point between an excitable state and an oscillatory state.

\end{document}